\documentclass[aps,prl,twocolumn,showpacs,superscriptaddress]{revtex4-2}
\usepackage{graphicx,amsmath,amssymb,amsfonts}
\usepackage[latin1]{inputenc}
\usepackage{array}
\usepackage{multirow}
\usepackage{float}
\usepackage{color}
\usepackage[normalem]{ulem}
\usepackage{booktabs}
\usepackage{siunitx}

\usepackage{hyperref}
\hypersetup{
	colorlinks = true,
	linkcolor = [rgb]{0.70, 0.13, 0.13},
	citecolor = [rgb]{0.25, 0.41, 0.88},
	urlcolor = [rgb]{0.25, 0.41, 0.88}
}
\usepackage{pifont}

\begin{document}

\title{Field-induced magnetic order in DyTa$_7$O$_{19}$ with two-dimensional pseudospin-$\frac{1}{2}$ triangular lattice}

\author{Feihao Pan}
\thanks{These authors contributed equally to this work}
\affiliation{Laboratory for Neutron Scattering and Key Laboratory of Quantum State Construction and Manipulation (Ministry of Education), School of Physics, Renmin University of China, Beijing 100872, China}

\author{Nan Zhao}
\thanks{These authors contributed equally to this work}
\affiliation{Institute of High Energy Physics, Chinese Academy of Sciences (CAS), Beijing 100049, China}
\affiliation{Spallation Neutron Source Science Center, Dongguan 523803, China}
\affiliation{Department of Physics, Southern University of Science and Technology, Shenzhen 518055, China}

\author{Songnan Sun}
\author{Chenglin Shang}
\affiliation{Laboratory for Neutron Scattering and Key Laboratory of Quantum State Construction and Manipulation (Ministry of Education), School of Physics, Renmin University of China, Beijing 100872, China}

\author{Peng Cheng}
\email[Corresponding author: ]{pcheng@ruc.edu.cn}
\affiliation{Laboratory for Neutron Scattering and Key Laboratory of Quantum State Construction and Manipulation (Ministry of Education), School of Physics, Renmin University of China, Beijing 100872, China}

\author{Jieming Sheng}
\email[Corresponding author: ]{shengjm@sustech.edu.cn}
\affiliation{School of Physical Sciences, Great Bay University and Great Bay Institute for Advanced Study, Dongguan 523000, China}

\author{Liusuo Wu}
\email[Corresponding author: ]{wuls@sustech.edu.cn}
\affiliation{Department of Physics, Southern University of Science and Technology, Shenzhen 518055, China}
\affiliation{Quantum Science Center of Guangdong-Hong Kong-Macao Greater Bay Area (Guangdong), Shenzhen 518045, China}
\affiliation{Shenzhen Key Laboratory of Advanced Quantum Functional Materials and Devices, Southern University of Science and Technology, Shenzhen 518055, China}

\begin{abstract}
	
The magnetic ground state of geometrically frustrated antiferromagnet attracts great research interests due to the possibility to realize novel quantum magnetic state such as a quantum spin liquid. Here we present a comprehensive magnetic characterization of DyTa$_7$O$_{19}$ with ideal two-dimensional triangular lattice. DyTa$_7$O$_{19}$ exhibits $c$-axis single-ion magnetic anisotropy. Although long-range magnetic order is not observed down to 100~mK under zero field, by applying a small magnetic field ($\sim$0.1~T), a magnetically ordered state with net magnetization of $M_s$/3 below $T_m$=0.14~K is identified ($M_s$ denotes the saturated magnetization). We argue that this state is an up-up-down magnetic structure phase driven by the dipole-dipole interactions between Ising-like spins of Dy$^{3+}$ in a two-dimensional triangular lattice, since its ordering temperature and temperature-field phase diagram can be well explained by the theoretical calculations based on dipolar interactions. DyTa$_7$O$_{19}$ could be viewed as a rare material platform that realizing pure Ising-like dipolar interaction in a geometrically frustrated lattice. 

\end{abstract}

\maketitle

\section{Introduction}
Materials with rare-earth based triangular-lattice have received a significant amount of attention in recent years. For one thing, the geometrical spin frustration and strong quantum fluctuations may lead to the novel quantum spin liquid state. Many candidate materials including YbMgGaO$_4$ and AYbX$_2$ (A=Cs,K,Na; X=S,O,Se) are proposed for this end\cite{YMGO, YMGO2, NRX27, NRX33, NRX35}. For another, they are also considered to be excellent adiabatic demagnetization refrigeration (ADR) coolants for achieving millikelvin temperature ranges. For examples, Na$_2$BaCo(PO$_4$)$_2$, KBaYb(BO$_3$)$_2$ and KBaGd(BO$_3$)$_2$ have shown exciting applications in sub-kelvin refrigeration\cite{ShengPNAS,Xiang2024,ADR,ADR2}. For whatever purpose it may serve, determining the magnetic ground state of a triangular-lattice antiferromagnet is essential for understanding the underlying physics.

In real magnets, the dipole-dipole interaction (DDI) always coexists with the exchange interaction. However, the dipolar interaction is often neglected because it is usually two to three orders of magnitudes lower than the exchange interaction. Especially in mangy transitional metal antiferromagnets, the exchange interaction strength could be above tens of Kelvin and that of DDI is usually below 0.1~K\cite{Cu1,Cu2}. While for rare-earth magnet with more localized $4f$ electrons which usually has very weak exchange interaction and large dipole moment, the DDI cannot be neglected and may play an important role in realizing various intriguing phenomena related to quantum magnetism. For example, Dy$_2$Ti$_2$O$_7$ with a highly frustrated pyrochlore lattice structure is known to host spin ice state in which magnetic monopoles may exist as emergent quasiparticles\cite{Ice1,Ice2}. This spin ice is often referred as a dipolar spin ice due to the dominant magnetic DDI in Dy$_2$Ti$_2$O$_7$. Besides, magnetic DDI is also proposed to be responsible for topological spin textures such as skyrmions\cite{batle_topological_2021, Skyrmions_Ezawa}. For materials with two-dimensional triangular lattice, although there were some theoretical proposals about magnetic ordered state or quantum spin liquid may exist considering mainly the DDI\cite{politi_dipolar_2006, keles_absence_2018}, the experimental reports are very rare.

The family of rare-earth heptatantalates RTa$_7$O$_{19}$ (R=$4f$ rare-earth ions) were recently reported to possess ideal two-dimensional triangular lattice of $R$ ions without any atomic-site disorder\cite{NM2022,TZM,Pan2024PRB}. Initially, no long-range magnetic order is detected for NdTa$_7$O$_{19}$ down to 40~mK and a spin-liquid like quantum-disordered ground state is proposed\cite{NM2022}. Later on, geometrically frustrated antiferromagnetic interactions were proposed in isostructural RTa$_7$O$_{19}$ (R=Ce, Pr, Dy, Ho, Yb) with magnetic measurements down to 2~K. However, as mentioned in our previous investigation on CeTa$_7$O$_{19}$ and YbTa$_7$O$_{19}$, the exchange interaction in RTa$_7$O$_{19}$ is extremely weak due to the large nearest $R-R$ distance ($\sim$6.2 \AA)\cite{Pan2024PRB}. Therefore any solid conclusions on the possible frustrated magnetism must be obtained with experimental characterizations at ultra-low temperatures. Moreover, the DDI should play an important role in this family of material due to the weak exchange interaction. The strongest DDI may be realized in DyTa$_7$O$_{19}$ which has the largest dipole moment. The previous magnetic characterization on DyTa$_7$O$_{19}$ was performed on polycrystalline samples and only with lowest accessible temperature of 2~K\cite{TZM}. Therefore experimental investigations on DyTa$_7$O$_{19}$ single crystals at lower temperatures are needed which may reveal critical information on the possible quantum magnetic state in RTa$_7$O$_{19}$.

In this study, magnetic susceptibility and specific heat measurements have been performed on single crystal of DyTa$_7$O$_{19}$ down to 70-100~mK. Its magnetic anisotropy is determined to be easy $c$-axis which may result from the crystal-electric-field effect. By applying a small magnetic field, an antiferromagnetic transition at 0.14~K is identified and its temperature-field phase diagram is obtained. All the experimental results are consistent with the DDI driven up-up-down magnetic order for Ising-like spins of Dy$^{3+}$ in a two-dimensional triangular lattice, which sheds new lights on the physical understanding of magnetism in RTa$_7$O$_{19}$ material family.

\section{methods}

Single crystals of DyTa$_{7}$O$_{19}$ were grown by chemical vapor transport reaction using NH$_4$Cl as the transport agent similar as that in previous publication\cite{DyTa7O19-CVT}. Firstly the precursor DyOCl powder was prepared by the solid-state reaction of NH$_4$Cl and Dy$_2$O$_3$, as in our previous work\cite{DyOCl}. Then DyOCl, Ta$_{2}$O$_{5}$ and Ta were mixed in a molar ratio of 2:1:1, ground and pressed into pellet. The mixture was sealed in an evacuated quartz tube and heated to \SI{950}{\celsius} at a rate of \SI{50}{\celsius}/h, then maintained at this temperature for 7 days. After furnace cooling to room temperature, the product was ground and mixed with additional amount of NH$_4$Cl, then sealed in an evacuated silica tube again. Typically, 15~mg NH$_4$Cl was used for a tube with inner diameter of 12~mm and length of 120~mm. For the last step, the mixture was heated at \SI{950}{\celsius} in a box-furnace for 7 days. Finally, the hexagonal plate-like DyTa$_{7}$O$_{19}$ single crystals with light yellow color and typical dimensions of 0.8$\times$0.8$\times$0.1 mm$^{3}$ could be found at the bottom of the tube.

X-ray diffraction (XRD) patterns of the samples were collected from a Bruker D8 Advance X-ray diffractometer using Cu K$_{\alpha}$ radiation. The elemental composition of single crystals were examined with energy dispersive x-ray spectroscopy (EDS, Oxford X-Max 50). For the magnetization measurements, a Quantum Design Magnetic Property Measurement System (MPMS) was used for temperatures above T=1.8~K and high-sensitive Hall sensor magnetometers were used for lower temperature down to 70~mK. The specific heat measurements were performed using the relaxation time method in a Quantum Design Physical Property Measurements System (PPMS). To reach the ultra-low temperature range (T=0.08{-}4.0~K), a dilution refrigerator insert was used.

\begin{figure}
	\includegraphics[width=7.8cm]{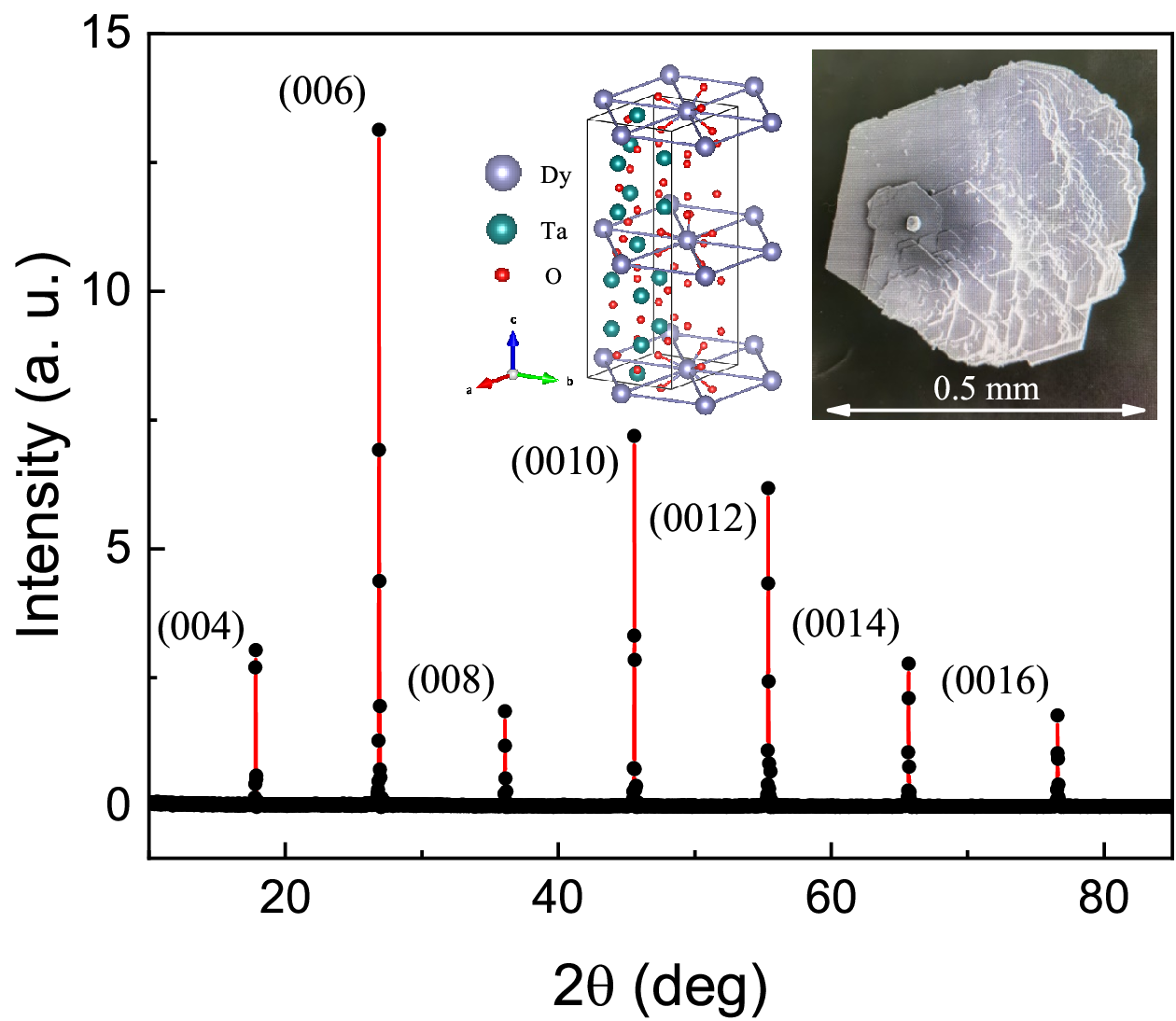}
	\caption {X-ray diffraction patterns from the $ab$-plane of DyTa$_7$O$_{19}$ single crystal. The inset shows its crystal structure and photo of one single crystal.} \label{Fig1}
\end{figure}

\section{Results and discussions}
\subsection{Crystal structure and magnetic anisotropy}

The crystal structure of DyTa$_7$O$_{19}$ features two-dimensional triangular lattice of Dy$^{3+}$ as illustrated in Fig.\ref{Fig1}. The nearest interplane Dy-Dy distance {9.939 \AA} is much larger than the nearest intraplane Dy-Dy distance {6.209 \AA}. Fig.\ref{Fig1} also shows a scanning electron microscope image of one DyTa$_7$O$_{19}$ single crystal, which demonstrates its layered structure with hexagonal size of the cleavage $ab$-plane. RTa$_7$O$_{19}$ is known as a material family without atomic site disorder. Previous powder and single crystal XRD analysis on DyTa$_7$O$_{19}$ also rule out any possible site mixing in this compound\cite{DyTa7O19-CVT,TZM}. Both the XRD measurement and EDS result confirm that our as-grown DyTa$_7$O$_{19}$ single crystal adopts the crystal structure and parameters same as previous reports\cite{DyTa7O19-CVT,TZM}.

\begin{figure*}[htbp]
	\centering
	\includegraphics[width=\textwidth]{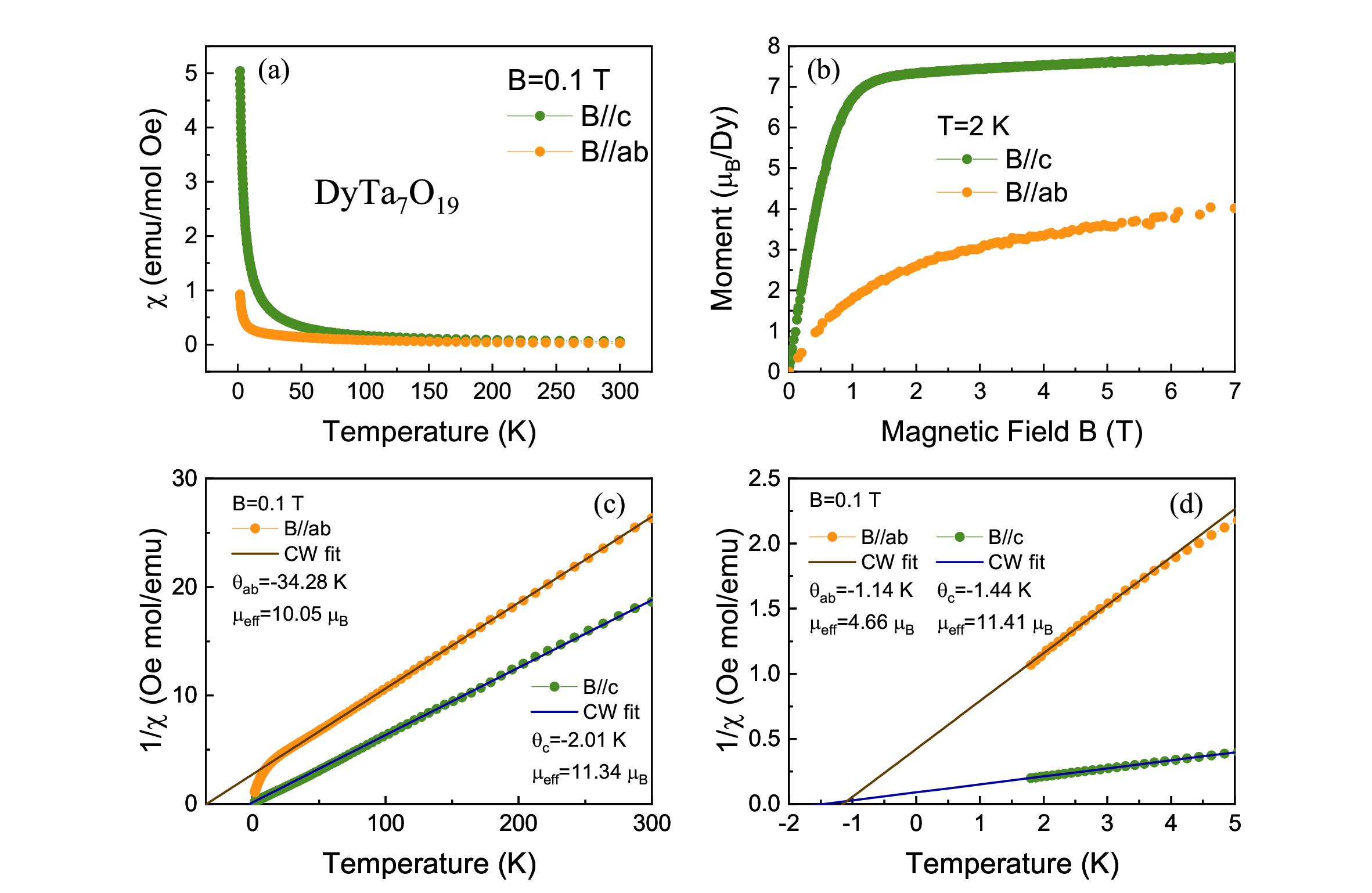}
	\caption{Magnetic characterization of DyTa$_7$O$_{19}$ above 1.8~K. (a) Temperature dependence of magnetic susceptibility with field along $c$-axis and parallel to the $ab$-plane. (b) Isothermal magnetization measured at 2~K with field applied along the two different directions. The results of Curie-Weiss fit on the inverse susceptibility at high- and low-temperature regimes are shown in (c) and (d) respectively.} \label{Fig2}
\end{figure*}

The results of magnetization measurements above 1.8~K are shown in Fig.\ref{Fig2}. From both the temperature-dependent susceptibility $\chi(T)$ and isothermal magnetization M(B) curves along different field directions, DyTa$_7$O$_{19}$ clearly exhibits a $c$-axis magnetic anisotropy. The saturation moment under B$\parallel$c $M_{sat}^{c}$ is 7.28 $\mu_B$/Dy at 2~K, which is much larger than $M_{sat}^{ab}$=2.44 $\mu_B$/Dy along B$\parallel$ab. This Ising-like anisotropy should be a rare-earth single-ion magnetic anisotropy determined by the crystal-electric-field (CEF) effect on the Dy ions. We have performed CEF calculations based on the point-charge model using both McPhase\cite{McPhase} and PyCrystalField\cite{PyCrystalField} packages. The derived CEF parameters yield single-ion anisotropy with an easy $c$-axis orientation, which is consistent with our experimental observations. The two lowest Kramers doublets of the Dy$^3+$ ion are determined as follows:
\begin{align*}
E_{0\pm} &= \pm 0.998|\pm 15/2\rangle + 0.07|\pm 9/2\rangle \pm 0.005|\pm 3/2\rangle
\end{align*}
with the first excited CEF level situated at an energy of 39.11 meV above the ground state. A notable feature emerges from the ground-state wavefunction composition: The contributions originate exclusively from the $|\pm 15 / 2\rangle$, $|\pm 9 / 2\rangle$ and $|\pm 3 / 2\rangle$ basis states. This particular admixture is constrained by the high local symmetry of the Dy$^{3+}$ site. Consequently, the minimum spin flip magnitude $\Delta S=3$ suggests that higher-order multipole interactions (e.g., quadrupole or octupole couplings) may appear beside conventional dipole-dipole interactions in the ultra-low temperature regime. This result may be considered as a rough reference for the CEF effect DyTa$_7$O$_{19}$ and needs to be further verified by the inelastic neutron scattering investigations.

\begin{figure*}[htbp]
	\centering
	\includegraphics[width=\textwidth]{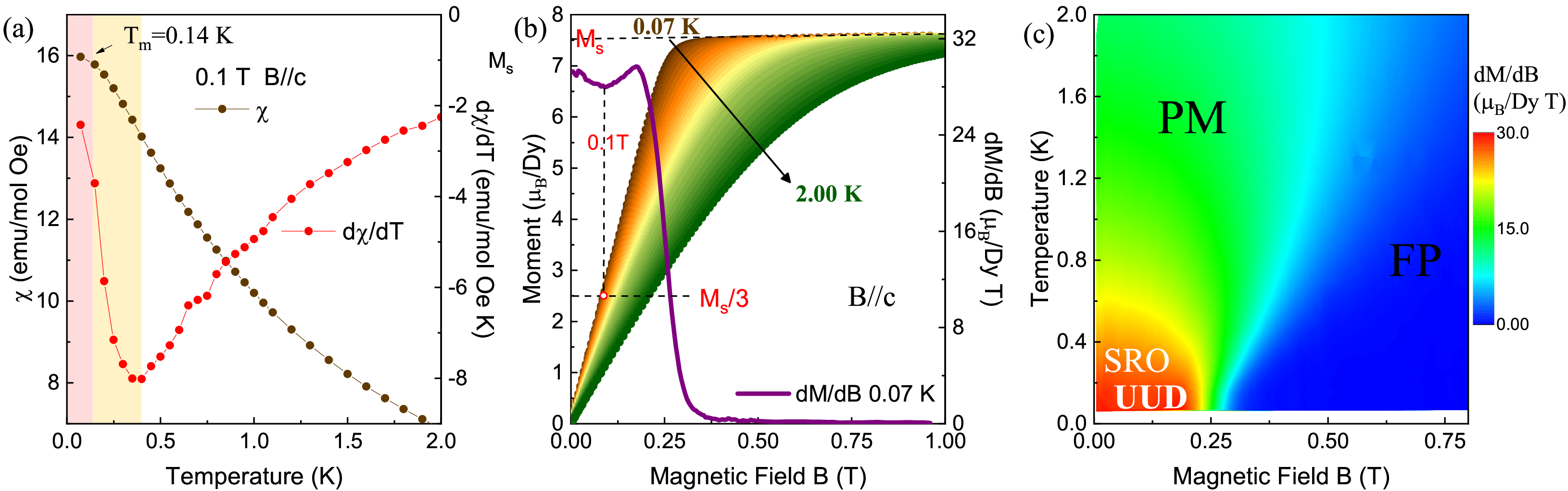}
	\caption {Magnetization measurement of DyTa$_7$O$_{19}$ down to 70~mK. (a) Temperature-dependent magnetic susceptibility $\chi(T)$ and $d\chi/dT$ in a field 0.1~T along $c$-axis. (b) Isothermal magnetization measured from 0.07~K to 2~K. A $dM/dB$ curve at 0.07~K shows a deep at 0.1~T where the net magnetic moment corresponds well with one-third of the saturation moment. (c) The contour plot of $dM/dB$ data in (b) which serves as the magnetic phase diagram of DyTa$_7$O$_{19}$.} \label{Fig3}
\end{figure*}

Fig.\ref{Fig2}(c) and (d) presents the results of Curie-Weiss (CW) fit on the inverse $\chi(T)$ data. The negative $\theta_{CW}$ temperatures is derived for all fitting regimes, which indicates dominate aniferromagnetic interactions in DyTa$_7$O$_{19}$. For rare-earth compounds with weak exchange interaction, The CW fit on the high-temperature region in Fig.\ref{Fig2}(c) yields effective moment close to the value of a free Dy$^{3+}$ ($\sim$10.6~$\mu_{B}$) and much larger $\theta_{CW}$ values than that obtained from fitting the low-temperature region in Fig.\ref{Fig2}(d). As in the previous studies on NdTa$_7$O$_{19}$\cite{NM2022} and Ce$_2$Zr$_2$O$_7$\cite{Gao2019}, for rare-earth magnets, the high-temperature CW fit may be affected by the crystal field effect and cannot truly reflect the strength of antiferromagnetic interaction, while the results from low-temperature CW fit could provide more reliable estimation. For DyTa$_7$O$_{19}$, the fit on the low-temperature regime shown in Fig.\ref{Fig2}(d) yields $\theta_{CW}^{ab}$=-1.14~K, $\mu_{eff}^{ab}$=4.66~$\mu_{B}$ for $B\parallel ab$ and $\theta_{CW}^{c}$=-1.44~K, $\mu_{eff}^{c}$=11.41~$\mu_{B}$ for $B\parallel c$. The value of $\mu_{eff}^{ab}$ has large deviation from the expectation of a free Dy$^{3+}$ and the calculated ground state wave function, which might be an unphysical result possibly due to stronger disturbance from crystal field effect along the hard-axis. While CW model can fit $\chi_{c}$ well and gives effective moment close to the expectation from the crystal field calculation. The obtained CW temperature indicates that the strength of antiferromagnetic interaction of DyTa$_7$O$_{19}$ may only be around 1~K, therefore magnetic characterization at ultra-low temperatures is needed to clarify its magnetic ground state.

\subsection{Field-induced magnetic order and phase diagram}

Fig.\ref{Fig3}(a) shows the magnetic susceptibility of DyTa$_7$O$_{19}$ with field of 0.1~T applied along $c$-axis down to 70~mK. There is a notable kink at $T_m$=0.14~K, which is likely due to an antiferromagnetic or ferrimagnetic transition. Later heat capacity data will confirm there is indeed a magnetic phase transition at $T_m$ under the same magnetic field. Moreover, there is a drastic change of susceptibility slope below 0.4~K as revealed from the $d\chi/dT$ curve in Fig.\ref{Fig3}(a), which possibly suggests short-range magnetic order may already develop at this temperature.

The isothermal magnetization data $M(B)$ below 2~K is shown in Fig.\ref{Fig3}(b). At the lowest temperature 70~mK, a magnetic field of 0.4~T along the $c$-axis is enough to polarize DyTa$_7$O$_{19}$ into a complete ferromagnetic state. A thorough measurement of $M(B)$ between 70~mK and 2~K allows us to make a contour plot of $dM/dB$ data, which is shown in Fig.\ref{Fig3}(c). This plot could be viewed as a rough magnetic phase diagram of DyTa$_7$O$_{19}$. The green area is the paramagnetic (PM) phase and the blue area is the fully polarized (FP) ferromagnetic phase. The red area consists of up-up-down (UUD) and possible short-range ordered (SRO) magnetic phases which will be explained later.

\begin{figure*}[htbp]
	\centering
	\includegraphics[width=\textwidth]{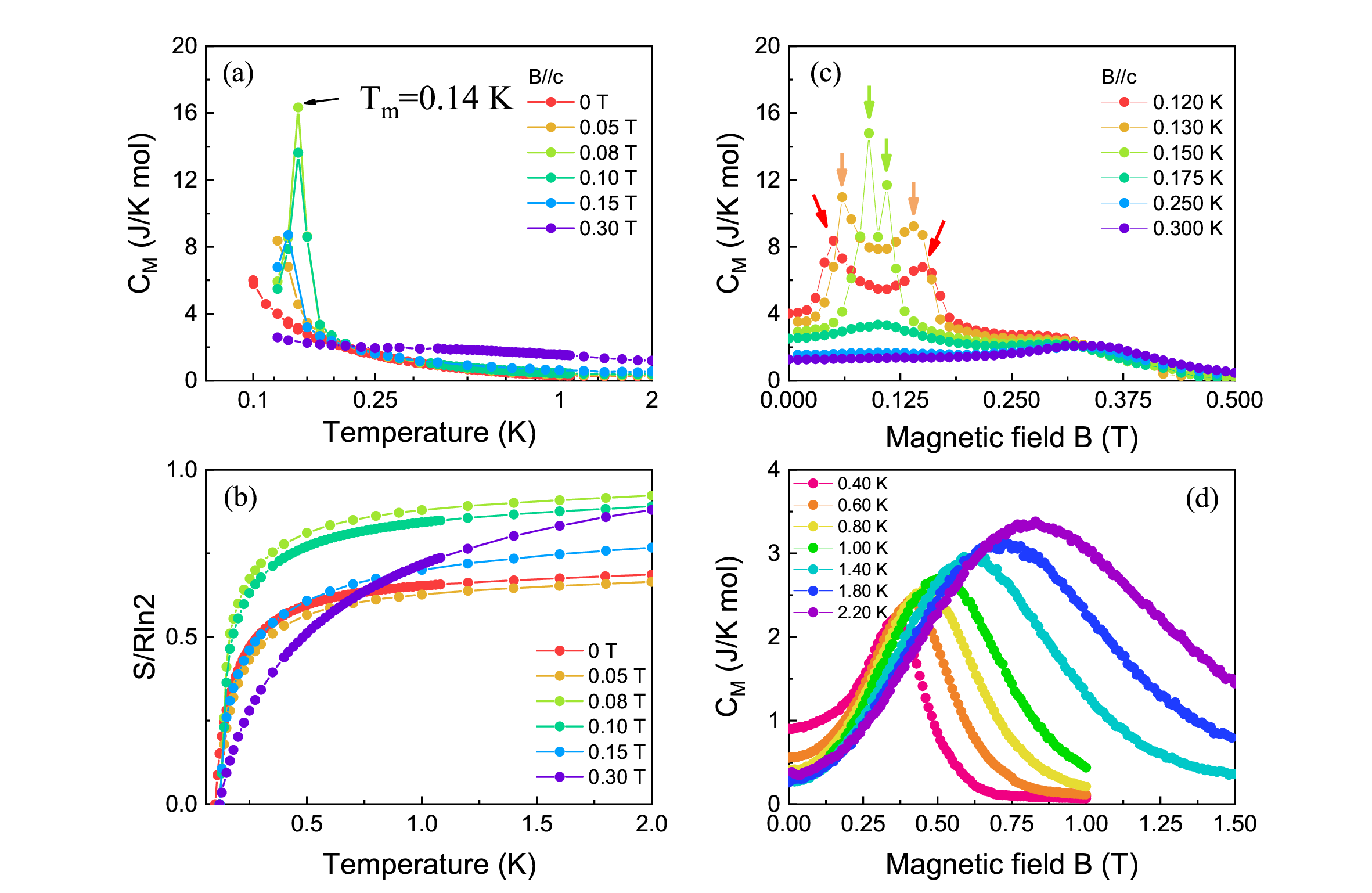}
	\caption {Heat capacity of DyTa$_7$O$_{19}$. (a) Temperature dependence of specific heat C$_M$ under different fields along $c$-axis. (b) Temperature-dependent magnetic entropy which is normalized to $Rln2$ and assuming a doublet ground state. (c,d) Magnetic field dependence of specific heat C$_M$ at different temperatures under B$\parallel$c.} \label{Fig4}
\end{figure*}

Heat capacity data in Fig.\ref{Fig4} at ultra-low temperatures provides critical information on the magnetic order in DyTa$_7$O$_{19}$. Below 2~K, the estimated phonon contribution to the specific heat data is negligibly small (below 0.1~J/K mol) as one can see from the heat capacity data of nonmagnetic YTa$_7$O$_{19}$ in a recent work\cite{SmTaO} and the raw $C_P(T)$ data could be viewed as the magnetic specific heat $C_M(T)$. Firstly, for the temperature dependent $C_M(T)$ without magnetic field, there is no sharp anomalies for temperatures down to 0.1~K, therefore any long-range magnetic ordered state is excluded above this temperature.

Secondly, $C_M(T)$ exhibits a sharp peak at $T_m$=0.14~K under field of 0.08~T and 0.1~T, indicating a magnetic phase transition to a long-range ordered state. The magnetization data in Fig.\ref{Fig3}(b) also provides crucial information in understanding this field-induced magnetic order. There is a deep-like anomaly at 0.1~T and 0.07~K (below $T_m$) in the $dM/dB$ curve. As marked in Fig.\ref{Fig3}(b), the net magnetic moment near this deep is exactly one-third of the saturation moment ($M_s$). For DyTa$_7$O$_{19}$ with two-dimensional triangular lattice and moments aligned along $c$-axis, the only physically reasonable magnetic structure for this field-induced order with net magnetization of $M_s/3$ should be an UUD phase. Additionally, the $\chi$(T) data under 0.1~T in Fig.\ref{Fig3}(a) exhibits a kink at $T_m$=0.14~K as in a UUD-like transition instead of a cusp as in a typical antiferromagnetic transition. In the next section, we will also show that the theoretical calculation also supports the existence of an UUD order.

\begin{figure*}[htbp]
	\centering
	\includegraphics[width=\textwidth]{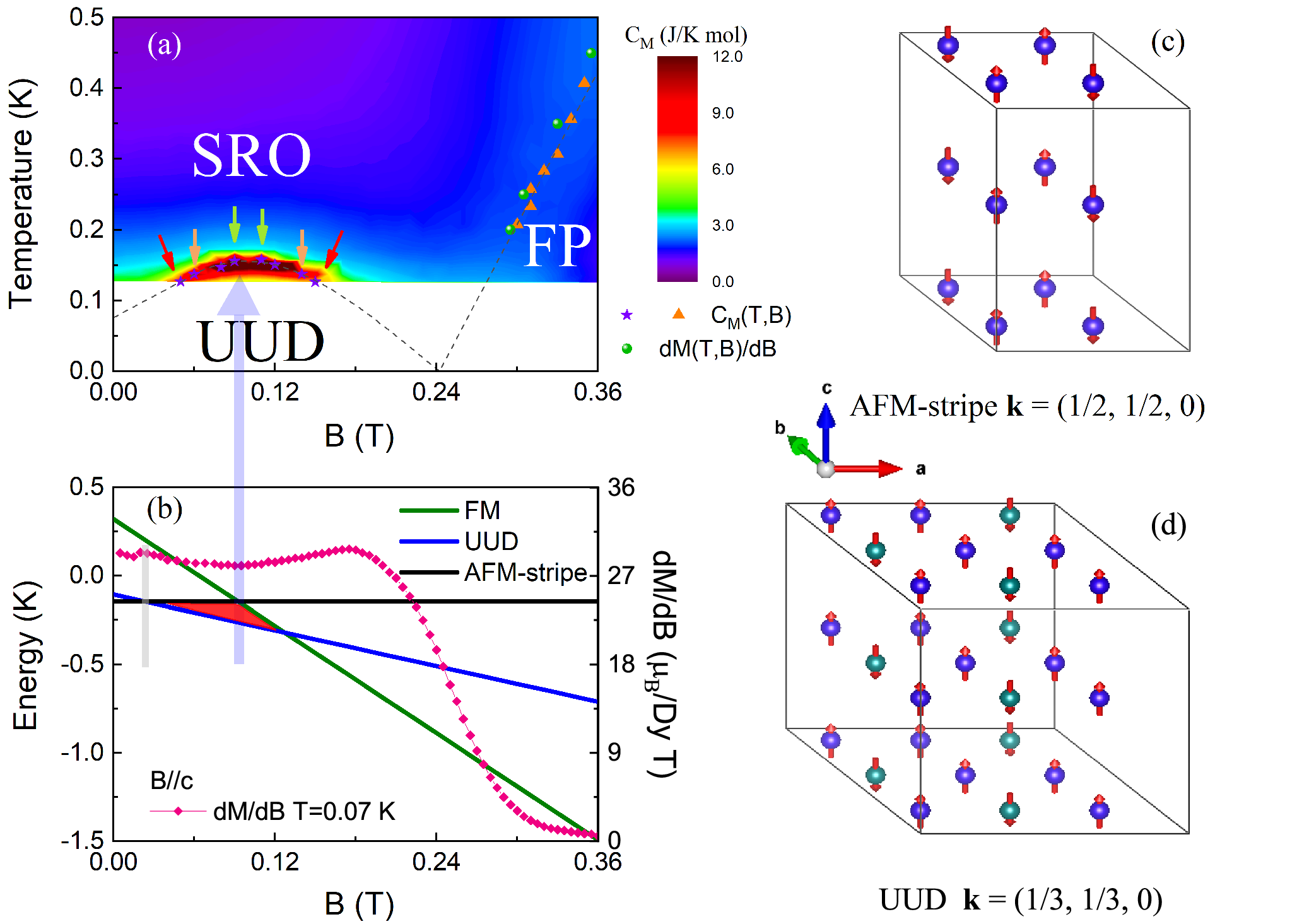}
	\caption {(a) Magnetic phase diagram of DyTa$_7$O$_{19}$ based on heat capacity data. (b) The calculated dipole-dipole energies as a function of field for different magnetic structures. The energy difference between UUD order and other magnetic orders is marked with red color at around 0.1~T. (c) Illustration of AFM-stripe magnetic structure with ferromagnetic interlayer coupling. (d) Illustration of UUD magnetic structure with ferromagnetic interlayer coupling.} \label{Fig5}
\end{figure*}

Interestingly, this field-induced UUD order shifts to lower temperatures with either increasing or decreasing magnetic field. This field-dependent behavior is also manifested in the double-peak feature in the isothermal field-dependent $C_M(B)$ data shown in Fig.\ref{Fig4}(c). The peaks are marked as five-pointed stars in the final phase diagram shown in Fig.\ref{Fig5}(a). The maximum $T_m$ determined from the field-dependent scan is $\sim$0.15~K which is close to that determined from temperature-dependent scan.

Next, with increasing magnetic field to above 0.3~T, a hump evolves in $C_M(B)$ as shown in Fig.\ref{Fig4}(d). This is a typical feature of Schottky anomaly that caused by the Zeeman splitting indicating a transition to the fully polarized ferromagnetic state\cite{zhao_TbScO_2023}. The center of these humps are marked using orange triangles in the phase diagram and corresponds well with the saturation field in the magnetization data [green spheres in Fig.\ref{Fig5}(a)].

In addition, the magnetic entropy was then calculated by integrating $C_M/T$ data and shown in Fig.\ref{Fig4}(b). Under field near 0.1~T, the total entropy saturates to about $Rln2$ at 2~K, consistent with the expectation of a pseudospin-$\frac{1}{2}$ Kramers doublet ground state for Dy$^{3+}$. We should mention that the magnetic entropy below the lowest accessible temperature is not included in this sum. Due to the instrument problem, the measurements under different magnetic fields do not all successfully reach the lowest temperature. Therefore the saturated entropy under zero field and 0.05~T has large deviation from $Rln2$, which implies large entropy releasing related to possible magnetic order should exist at temperatures lower than 0.1~K.

Finally, the above experimental results will allow us to present the magnetic phase diagram of DyTa$_7$O$_{19}$. Firstly, the contour plot of $dM/dB$ in Fig.\ref{Fig3}(c) clearly identifies a border between paramagnetic and ordered state in the lower left corner with temperatures which are higher than $T_m$ determined from heat capacity. A likely explanation for this feature from above $T_m$ to the bordering temperatures is a precursor short-ranged order before the final formation of long-range UUD order. This explanation is also supported by the slope change of $d\chi/dT$ [Fig.\ref{Fig3}(a)]. Furthermore, a detailed phase diagram based on magnetization and specific heat data is also plotted in Fig.\ref{Fig5}(a). The UUD and FP regions are determined from both magnetization and specific heat data discussed above. Next, we argue the field-induced long-range UUD magnetic order below $T_m$ should be driven by magnetic dipolar interactions based on the theoretical calculations given in the next section.

\subsection{Magnetic dipolar interactions}

The large distance between neighboring Dy ions, the localized $4f$ electrons and the above experimental facts all indicate a weak exchange interaction in DyTa$_7$O$_{19}$. Therefore the magnetic dipole-dipole interaction (DDI) may play an important role in determining the magnetic ground state of DyTa$_7$O$_{19}$ due to the large local moment of Dy$^{3+}$. Next, we present the theoretical calculations on the dipole-dipole energies for different magnetic structures of DyTa$_7$O$_{19}$. Since DyTa$_7$O$_{19}$ exhibits a strong easy-axis magnetic anisotropy, we fixed the moments of Dy$^{3+}$ along the $c$-axis. Then the ferromagnetic (FM), antiferromagnetic-stripe (AFM-stripe) and UUD type structures in typical two-dimensional triangular lattice with either interlayer ferromagnetic or antiferromagnetic coupling were considered. The DDI energy for different magnetic structures can be expressed as
\begin{equation*}
    E_{\text {dipole}}=-\frac{\mu_{0}}{4 \pi} \sum_{\mathrm{i<j}} \frac{1}{\left|\mathbf{r}_{\mathrm{ij}}\right|^{3}}\left[3\left(\mathbf{m}_{\mathrm{i}} \cdot \hat{\mathbf{r}}_{\mathrm{ij}}\right)\left(\mathbf{m}_{\mathrm{j}} \cdot \hat{\mathbf{r}}_{\mathrm{ij}}\right)-\left(\mathbf{m}_{\mathrm{i}} \cdot \mathbf{m}_{\mathrm{j}}\right)\right]
\end{equation*}
where \({\mu_{0}}\) is the vacuum permeability, \(\mathbf{r}_{\mathrm{ij}}\) is the distance vector between local moment \(\mathbf{m}_{\mathrm{i}}\) and \(\mathbf{m}_{\mathrm{j}}\), and \(\hat{\mathbf{r}}_{\mathrm{ij}}\) is the unit vector in the direction of \(\mathbf{r}_{\mathrm{ij}}\).
Although DDI is inherently long-range, we restrict the calculation to interactions within a \(2\times2\) or \(3\times3\) magnetic cell. Interactions with atoms beyond this range have little influence due to the \(\mathrm{r}^{\mathrm{-3}}\) decay.

In the absence of magnetic field, the AFM-stripe structure with with propagation vector \(\mathbf{k}=(1/2, 1/2, 0)\) shown in Fig.\ref{Fig5}(c) has the lowest average dipole-dipole energy per atom which is -0.147~K. When a magnetic field is applied, the total energy is given by \(E=E_{\text {dipole}}+E_{\text {Zeeman}}\), where Zeeman energy is defined as
\begin{equation*}
E_{\text {Zeeman}}=-\sum_{\mathrm{i}} \mathbf{B} \cdot \mathbf{m}_{\mathrm{i}}
\end{equation*}
The calculation result shows that the total energy of the AFM-stripe phase remain unchanged under magnetic field. However the energy of an UUD magnetic structure with \(\mathbf{k}=(1/3, 1/3, 0)\) shown in Fig.\ref{Fig5}(d) becomes the lowest one when magnetic field is larger than $\sim$0.02~T. When the field is larger than 0.12~T, the FM phase eventually becomes the most energy favorable magnetic ordered state. The field-dependent total energies of the above three magnetic phases are shown specifically in Fig.\ref{Fig5}(b). The calculation result shows that other magnetic structures with interlayer antiferromagnetic coupling are not favorable under any magnetic field, which is consistent with the common expectation for pure dipolar interaction and the related result is not shown here.

The above calculation result has demonstrated close connections to the experimentally determined magnetic phase diagram of DyTa$_7$O$_{19}$ in Fig.\ref{Fig5}(a). Theoretically at around 0.1~T, the DDI driven UUD phase is most energy-favorable and has the largest energy difference comparing with FM and AFM-stripe phases as one can see from Fig.\ref{Fig5}(b). Then experimentally, the magnetic order becomes most prominent under the same magnetic field and the ordering temperature decreases when deviating from this field. This consistency not only provide additional evidence for the existence of the UUD order but also strongly suggests the UUD order is driven by DDI. Additionally, for the energy crossover between AFM-stripe and UUD phases at $\sim$0.02~T, we notice that there is also an anomaly detected in the $dM/dB$ curve under the same field [Fig.\ref{Fig5}(a)]. When applied field is larger than 0.2~T, the expected FM phase from DDI calculation is also observed experimentally. Therefore, the above results indicate the magnetic order in DyTa$_7$O$_{19}$ is driven by magnetic dipolar interactions.

Under zero magnetic field, the calculated dipole-dipole energy of AFM-stripe phase is -0.147~K. However no long range order is observed down to 0.1~K, which implies that magnetic frustration of either DDI or exchange interaction might exist in DyTa$_7$O$_{19}$. A spin liquid ground state is possible but no solid conclusion can be drawn based on the current work. Under magnetic field of $\sim$0.1~T, the ordering temperature $T_m$=0.14~K of the UUD phase is comparable to the calculated dipole-dipole energy 0.26~K. The experimentally observed dome-like evolution of $T_m$ with increasing field in Fig.\ref{Fig5}(a) is most likely caused by the energy competition between UUD and AFM-stripe/FM magnetic phases as shown in Fig.\ref{Fig5}(b). In addition, it might also be partially explained by the suppression of magnetic frustration with increasing magnetic field.

\subsection{Discussion}

Recent investigations on CeTa$_7$O$_{19}$\cite{arx} and SmTa$_7$O$_{19}$\cite{SmTaO} have confirmed the absence of any long-range magnetic order down to $\sim$30~mK\cite{arx,SmTaO}. $\mu$SR, specific heat and thermal conductivity suggest possible quantum spin liquid state in these two compounds. For isostructural DyTa$_7$O$_{19}$, we have discovered a field-induced long-range UUD magnetic order existing below 0.14~K, which has never been realized before in the RTa$_7$O$_{19}$ material family.

The UUD structure in triangular-lattice antiferromagnets has been long-predicted and drawn great attentions in the research of quantum magnetism. It has often been observed as a field-induced one-third magnetization plateau and explained as arising from an order-by-disorder mechanism driven by quantum fluctuations\cite{p1,p2,p6,NP2019}. Especially, recent theoretical work has pointed out that in the phase diagram of the quantum Ising model on a triangular lattice under external field, the quantum phase transitions between the UUD and other phases could exist\cite{Meng}. In this work, we have proposed another model for explaining the field-induced magnetic order in the triangular-lattice antiferromagnet DyTa$_7$O$_{19}$, that is a classical physical picture based on magnetic dipolar interaction. Our findings give insights on understanding the elusive magnetism in quantum magnets with weak exchange interactions.

On the other hand, the weak exchange interaction, low ordering temperature and large magnetic moment may make DyTa$_7$O$_{19}$ potentially useful for adiabatic demagnetization refrigeration.  As magnetic moments are easily aligned by the external magnetic field, causing a reduction of entropy. Especially, DyTa$_7$O$_{19}$ has very small saturation field ($\sim$0.4 T below 1 K), which may allow ADR application using simple and cheap permanent magnet.

\section{Conclusions}

To summarize, we have synthesized the single crystals of DyTa$_7$O$_{19}$ with geometrically frustrated triangular-lattice. Its easy $c$-axis magnetic anisotropy has been determined through anisotropic magnetization measurements. Calculation and measurement of magnetic entropy confirms the pseudospin-$\frac{1}{2}$ Kramer doublet ground state. A field-induced UUD magnetic order at $T_m$$\sim$0.14~K is observed from both ultra-low-temperature magnetic susceptibility and specific heat measurements. The evolution of this UUD phase under magnetic field agrees well with our theoretical calculations based on the magnetic dipole-dipole interactions in DyTa$_7$O$_{19}$. We propose that DyTa$_7$O$_{19}$ represents a material platform for investigating Ising-like dipolar interactions in a geometrically frustrated lattice.

\section*{Acknowledgement}
This work was supported by the National Natural Science Foundation of China (No. 12074426, No. 12474148 and No. 11974157), the National Key Research and Development Program of China (Grant No. 2021YFA1400400), the Stable Support Plan Program of Shenzhen Natural Science Fund (Grant No. 20231121101954003), the Open Fund of the China Spallation Neutron Source Songshan Lake Science City (Grant No. KFKT2023A06) and Guangdong Provincial Quantum Science Strategic Initiative (Grant No. GDZX2401007).

\bibliography{DyTaO}{}
\end{document}